\documentclass[traditabstract]{aa} 
\usepackage{graphicx}
\usepackage{longtable}
\usepackage{rotating}
\usepackage{natbib}
\usepackage[usenames]{color}
\usepackage{multirow}
\usepackage{url}
\usepackage{ulem}
\usepackage{supertabular}
\usepackage{aalongtable}
\usepackage{afterpage}
\bibpunct{(}{)}{;}{a}{}{,} 
\newcommand{\ratio} {N({\rm H}_2) / I_{\rm CO(1-0)}}
\newcommand{\ratioo} {N({\rm H}_2) / I_{\rm CO}}

\newcommand{\kms}   {{\rm \  km \  s^{-1}}}

\newcommand{\NHmol} {N(\mathrm{H_{2}})}
\newcommand{\X} {\NHmol/\mathrm{I}_{\mathrm{CO}}}
\newcommand{\Htwo} {\rm H_{2}}
\newcommand{\Xunit} {\,{\rm cm^{-2}/(K\kms)}}

\usepackage{txfonts}
\begin{document}
   \title{Molecular Cloud Formation and the Star Formation Efficiency in M~33}
   \titlerunning{Molecule and Star Formation in M~33}
   \subtitle{Molecule and Star Formation in M~33}

   \author{J. Braine \inst{1} \and P. Gratier \inst{1} \and C. Kramer  \inst{2}   \and K.F. Schuster  \inst{3} \and F. Tabatabaei \inst{4}  \and E. Gardan \inst{1}         }

   \institute{Laboratoire d'Astrophysique de Bordeaux, Universit\'e de Bordeaux, OASU, CNRS/INSU, 33271 Floirac France\\
              \email{braine@obs.u-bordeaux1.fr}
         \and
            IRAM, 300 Rue de la piscine, F-38406 St Martin d'H\`eres, France   
         \and
            IRAM, Avenida Divina Pastora, 7, Nucleo Central E 18012 Granada, Spain   
         \and
            Max-Planck Institut fŸr Radioastronomie, Auf dem H\"ugel 69, 53121 Bonn, Germany
        }

   \date{}

  \abstract{Does star formation proceed in the same way in large spirals such as the Milky Way and in smaller chemically younger galaxies?  Earlier work suggests a more rapid transformation of H$_2$ into stars in these objects but (1) a doubt remains about the validity of the H$_2$ mass estimates and (2) there is currently no explanation for why star formation should be more efficient.  M~33, a local group spiral with a mass $\sim 10$\% and a metallicity half that of the Galaxy, represents a first step towards the metal poor Dwarf Galaxies.    We have searched for molecular clouds in the outer disk of M~33 and present here a set of detections of both $^{12}$CO and $^{13}$CO, including the only detections (for both lines) beyond the R$_{25}$ radius in a subsolar metallicity galaxy.  The spatial resolution enables mass estimates for the clouds and thus a measure of the $\ratioo$ ratio, which in turn enables a more reliable calculation of the H$_2$ mass.  Our estimate for the outer disk of M~33 is $\ratio \sim 5 \times 10^{20} \Xunit$ with an estimated uncertainty of a factor $\le 2$.  While the $^{12/13}$CO line ratios do not provide a reliable measure of $\ratioo$, the values we find are slightly greater than Galactic and corroborate a somewhat higher $\ratioo$ value.  Comparing the 
CO observations with other tracers of the interstellar medium,  
no reliable means of predicting where CO would be detected was identified.  In particular, CO detections were often not directly on local HI or FIR or H$\alpha$ peaks, although generally in regions with FIR emission and high HI column density.  
The results presented here provide support for the quicker transformation of H$_2$ into stars in M~33 than in large local universe spirals. }

\maketitle
\section{Introduction}

Several recent papers \citep{Leroy06, Gardan07, Gratier10} have suggested that the rate of transformation of molecular gas (H$_2$) into stars is higher in small chemically young galaxies -- those with low metallicities.
The latter articles devoted much attention to whether the conversion factor from CO to 
$\Htwo$, $\X$, usually expressed in H$_2$ molecules $\Xunit$, could be severely underestimated.  
In this article, we present sensitive measurements of the CO emission from the outer disk of M~33, allowing us to estimate Virial masses for isolated molecular clouds far from the center as well as some $^{13}$CO line measurements, providing a further check on the physical conditions in the Giant Molecular Clouds (GMCs) in M~33.  A further goal is naturally to understand the mechanisms of molecular cloud formation and the outer disk provides a means to explore physical conditions unexplored by earlier work which only presented molecular cloud data in the inner disk where the stellar mass surface density dominates that of the gas.

\begin{figure*}[!h]
\begin{flushleft}
\includegraphics[angle=0,width=18cm]{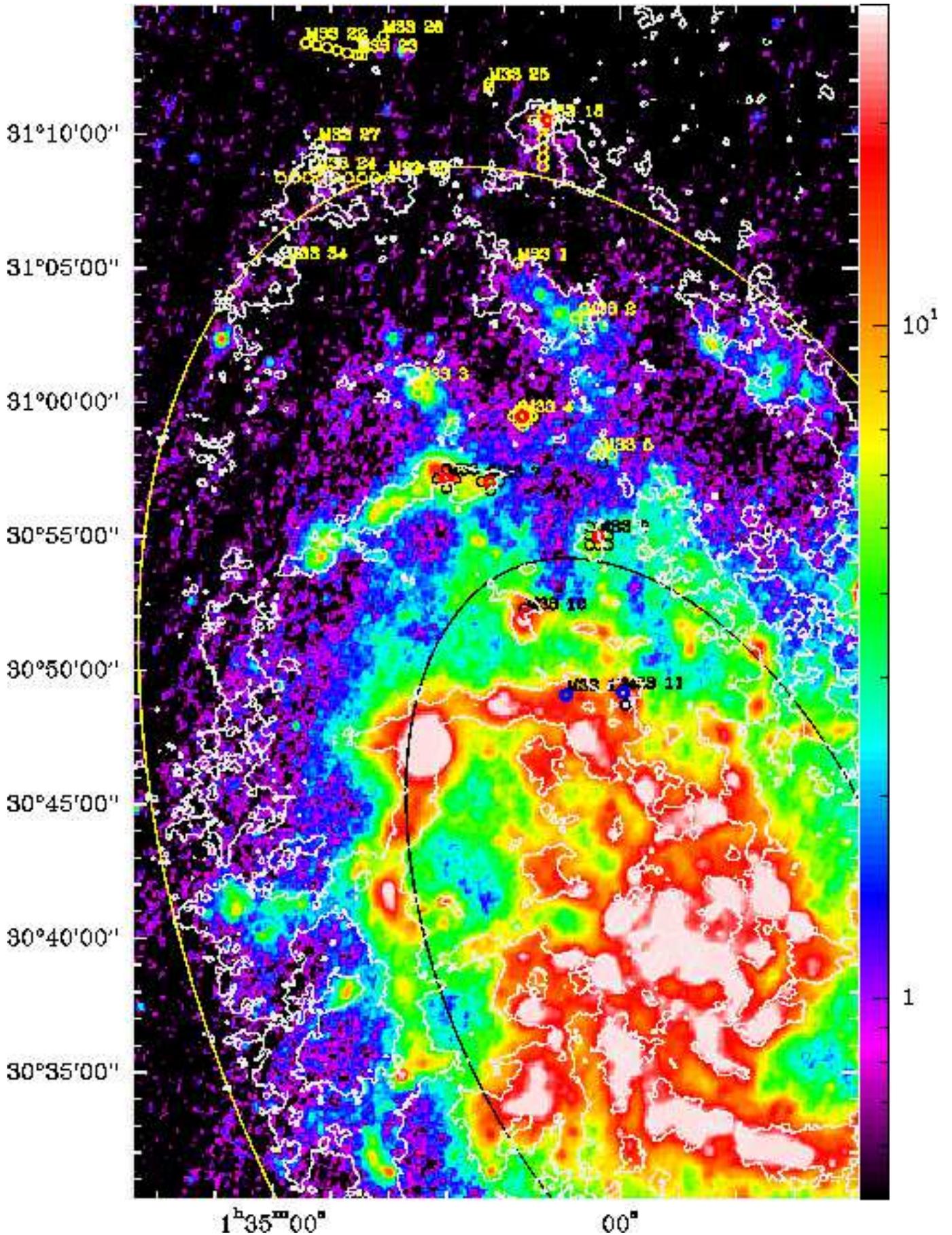}
\caption{ 70$\mu$m image of M~33 \citep{Tabatabaei07a} using a logarithmic scale to show the outermost features.  The R$_{25}$ radius is illustrated with a yellow ellipse.  The inner black ellipse indicates half the R$_{25}$ radius, where gaseous and stellar surface densities are roughly equal.  The positions observed in $^{12}$CO are shown in yellow in the outer parts and in black in the inner parts.  Positions observed in $^{13}$CO are shown in red except for the two innermost points (M33\_11 and M33\_12) which are shown in blue in order to be visible.  All the positions observed in  $^{13}$CO have also been measured in $^{12}$CO.
The white contour shows the $1.5 \times 10^{21}$ cm$^{-2}$ HI contour at 17$''$ resolution from \citet{Gratier10b}. }
\end{flushleft}
\end{figure*}

In particular, we present data on a series of mid to outer disk clouds in M~33, including an interarm GMC with no measured associated star formation (H$\alpha$ or Far-IR emission) and a GMC beyond the R$_{25}$ radius.  Because the stellar population in M~33 is rather young, the R$_{25}$ radius \citep[30.8$'$][]{Paturel03}, defined in B band, corresponds to an extremely low stellar surface mass density, well below that of the gas at the same radius.  In M~33, the radius at which stellar and gaseous surface densities are equal is about 0.5 R$_{25}$ whereas in large spirals the stars dominate until about the R$_{25}$ radius.

It has long been known that the amount of CO emission per unit star formation or per unit H$_2$ is
lower, sometimes much lower, in galaxies with subsolar metallicity \citep[e.g.][]{Rubio91}.
In the solar neighborhood, the conversion factor $\X$ is of order $2 \times 10^{20} \Xunit$ with a strong increase from the center to the outer disk \citep[e.g.][]{Sodroski95}.  This radial variation is not only true for the Galaxy but also for other spirals where it has been studied \citep{Braine_n4414b, Braine_n3079}.  The amount of CO emission per H$_2$ mass varies not only with metallicity but also with the radiation field, which can act to raise the CO emission by warming the gas.  If the metallicity is low the effect is opposite because the lower dust content and lesser degree of self-shielding by molecules other than H$_2$ reduce the size of the CO-emitting regions with respect to the H$_2$ cloud size, such that the CO emission principally traces the central regions of molecular clouds.  Because the CO lines are optically thick, moderate changes in metallicity and radiation field with respect to Galactic conditions (solar vicinity or molecular ring) can be expected to only generate moderate changes in the $\X$ ratio.

\clearpage

The standard techniques to estimate the H$_2$ mass, and thus $\X$ ratio, are through the dust emission and an assumed dust-to-gas mass ratio, the "virial" mass from the cloud size and line width, and from optically thin tracers such as $^{13}$CO.  We discuss the latter two methods in Sections 4.1 and 4.2.  The resulting $\ratioo$ is then used to estimate the star formation efficiency (SFE), defined as the ratio of the star formation rate per unit H$_2$, in M~33.

M~33 has a metallicity a factor 2 below solar \citep{Rosolowsky08, Magrini09}, a moderate UV radiation field, and is a spiral galaxy despite having a mass $\sim 10$ times lower than the Galaxy.  It thus represents a first step towards the still smaller, lower metallicity, irregular galaxies in the Local Group and beyond.  
Based on the above, we assume an Oxygen abundance of $12 + log(O/H) = 8.4-0.03 R_{kpc}$.  
The distance to M~33 is assumed to be 840 kpc and we
adopt an inclination angle of 56$^\circ$ and a position angle of 22.5$^\circ$ as in \citet{Gardan07}.

\section{Observations}

The observations presented here are a follow-up to the \citet{Gardan07} mapping of a large part of M33.
All data were taken with the 30meter telescope run by the Institut de RadioAstronomie Millim\'etrique (IRAM) on Pico Veleta near Granada, Spain.  Observing runs were in April, August, and Nov. of 2006.  

The $^{12}$CO and $^{13}$CO $J=1\rightarrow 0$ and $J=2\rightarrow 1$ transitions were observed with the "AB" receivers.  These receivers are dual polarization and the beam-splitter enables simultaneous observation of the two transitions.
Data were taken under good conditions, with system temperatures (T$_a^*$) of 300 -- 400 K except for the $^{13}$CO(1--0) where the average system temperature was $\sim 200$K.  As a result, data reduction was simple: bad channels were eliminated, a zero-order baseline (constant value) was subtracted, and spectra were averaged for each position.

The positions observed are shown as yellow and black circles in Fig. 1 for the $^{12}$CO and red or blue circles for the $^{13}$CO.  The size illustrates the IRAM CO(1--0) half-power beamsize.  
Generally, both the $J=1\rightarrow 0$ and $J=2 \rightarrow 1$ transitions were observed simultaneously.
The data are presented using the main beam temperature scale.  The main beam efficiencies are estimated to be 
0.72, 0.74, 0.53, and 0.54 for respectively the lines at 115, 110, 230, and 220 GHz (see {http://www.iram.es/IRAMES/mainWiki/Iram30mEfficiencies}) and forward efficiencies 0.95, 0.95, 0.91, and 0.91.

Table 1 provides the positions at which we observed $^{12}$CO.
Tables 2 and 3 give the $^{12}$CO $J=1\rightarrow 0$ and $J=2\rightarrow 1$ fluxes at each detected position and
the $rms$ noise level at 1MHz resolution for the undetected positions.  The results of fitting gaussian line profiles 
to the spectra, in order to robustly measure line widths and velocities, can be found in Table 4.
Table 5 gives the  $^{13}$CO fluxes and $^{12/13}$CO line ratios for both transitions.

\section{Molecular clouds in the outer disk of M~33}

\begin{figure*}[!h]
\begin{flushleft}
\includegraphics[angle=270,width=18cm]{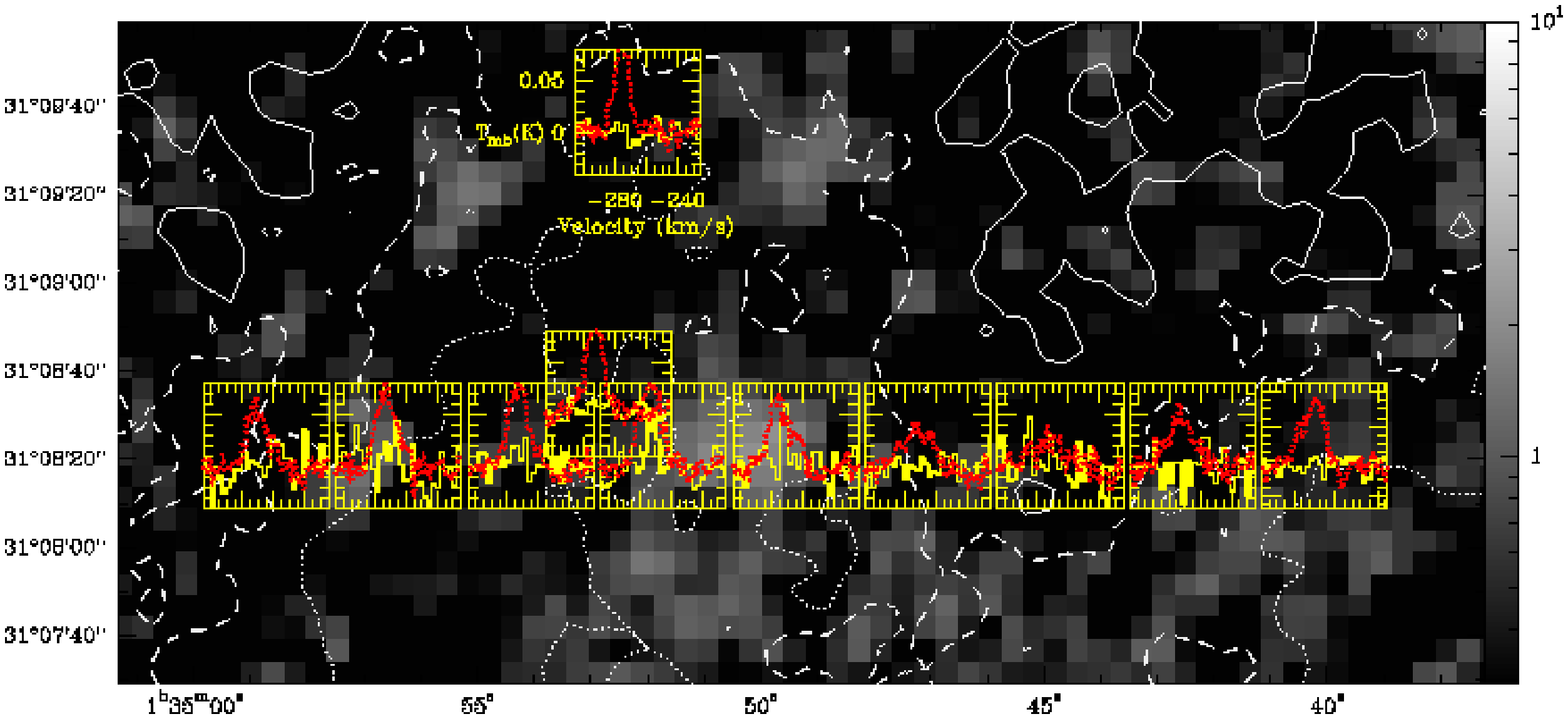}
\caption{ CO(1--0) spectra (yellow) and HI spectra (red dotted) of sources M33\_20, M33\_24, and M33\_27 
on 70$\mu m$ image with HI contours at 0.8, 1.2 (dashed), and 1.6 $\times 10^{21}$ cm$^{-2}$ (dotted).  
The wedge shows the 70$\mu m$ image scale in MJy sr$^{-1}$.  Note that two positions along the strip (M33\_20) have been 
detected but that they correspond to neither HI nor 70$\mu m$ maxima.}
\end{flushleft}
\end{figure*}

\begin{figure*}[!h]
\begin{flushleft}
\includegraphics[angle=270,width=18cm]{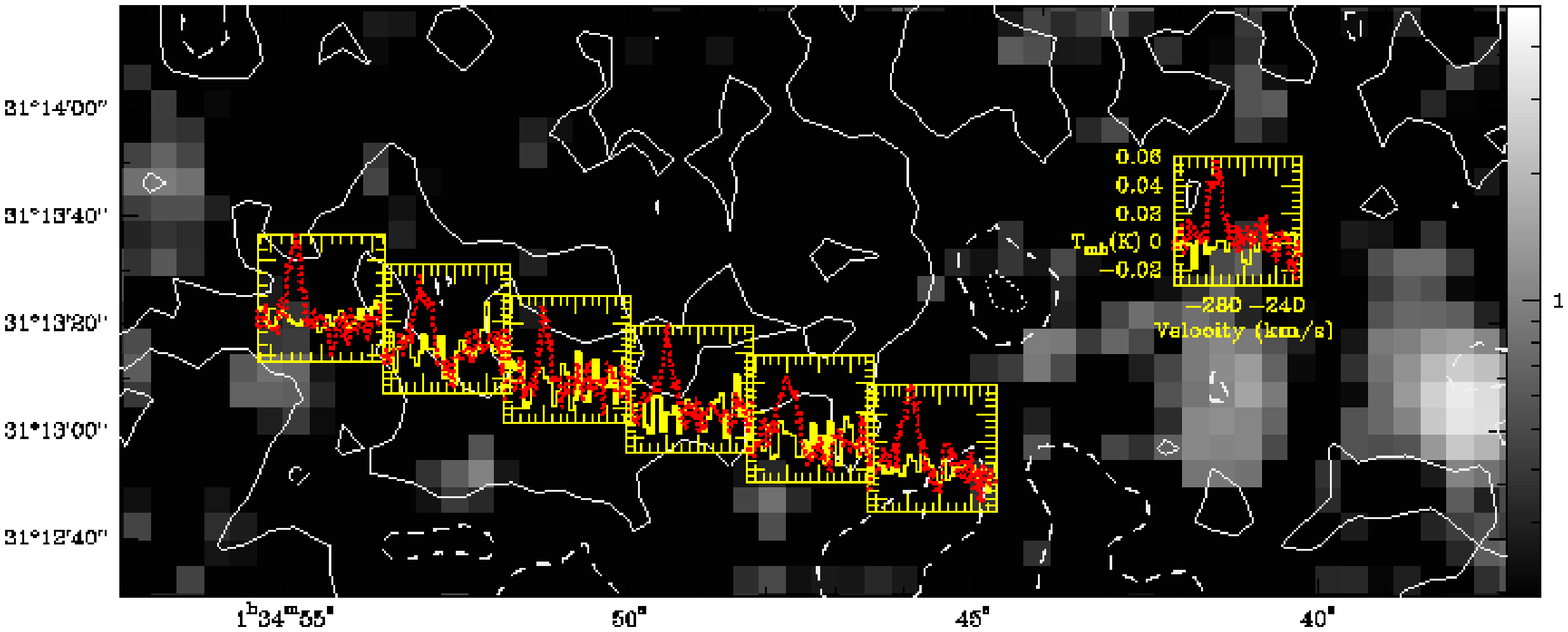}
\caption{ CO(1--0) spectra (yellow) and HI spectra (red dotted) of sources M33\_22, M33\_23, and M33\_26 
on 70$\mu m$ image with HI contours at 0.8, 1.2 (dashed), and 1.6 $\times 10^{21}$ cm$^{-2}$ (dotted).  
The wedge shows the 70$\mu m$ image scale in MJy sr$^{-1}$.  None of the positions have been detected in CO.}
\end{flushleft}
\end{figure*}

In this article we present CO and even $^{13}$CO detections beyond the R$_{25}$ radius of M~33.
These are the first such detections for a sub-solar metallicity galaxy.
Only a few detections have been made in the outer parts of galaxies: in the Milky Way, using the 
rotation curve to estimate the distance \citep[e.g.][]{Brand87,Digel94}, in NGC~4414 \citep{Braine04b}, and in NGC~6946 \citep{Braine07}.  One of the issues is whether large quantities of molecular gas could have escaped detection in the outer disk \citep{Pfenniger94a}.  In very nearby galaxies such as M33, molecular clouds can be spatially resolved, allowing the use of the Virial theorem and isotopic lines to (roughly) assess their masses and 
resemblance to inner disk clouds.  This is particularly interesting for small, chemically young, galaxies like M~33. 

After the mapping of the northern part of M~33 by \citet{Gardan07}, we wanted to integrate more deeply at positions 
where we thought CO might be found, with the goal of testing whether we could predict from other means if molecular gas were present.  As molecular clouds form out of the atomic gas, which dominates in the outer disk, one of the criteria was the presence of high HI column density.  Dust emission is frequently linked to CO emission but in the 
outer parts of M~33 the FIR and MIR emission was often too weak to be detected by Spitzer (or earlier satellites), 
despite the substantial HI column densities.  A series of positions was observed to a lower noise level, and in CO
(1--0), rather than in the CO(2--1) mapping by \citet{Gardan07}; the coordinates are given in Tab. 1 and they are shown in
Fig. 1 as circles.  Figures 2, 3, and 4 present the CO(1--0) and HI spectra at these positions on the 70$\mu m$ 
image with HI column density contours.  The HI data is at 17$"$ resolution from a mosaic of VLA C and D configuration fields described in \citet{Gratier10b}.  These outer positions were chosen because either star formation was present
from the H$\alpha$ image \citep{Greenawalt98,Hoopes00} or high HI antenna temperatures (either narrow lines or 
high total column densities) were found in the
\citet{Deul87} data.  The FIR data was not available at the time of the observations yet the spectra show that the
CO detections are not always at the HI or FIR/H$\alpha$  maxima although they tend to be part of larger HI structures with 
H$\alpha$ emission.  At higher sensitivity, the FIR emission would probably be detected as well.  Here we show the 
70$\mu m$ data but the 160$\mu m$ images do not show more extended emission. 
Figures 2, 3, and 4 show the sources labelled M~33 18, 20, 22, 23, 24, 26, and 27 -- all of which are north of Dec. 31:05.  Only deep NIR images detect the stars at these distances from the center.

Fig. 5 shows the CO(1--0) and CO(2--1) spectra of the outer disk clouds M33\_2, M33\_3, M33\_5, and the CO(1--0) spectra of the two detected positions in M33\_20.  
Fig. 6 shows the 8 clouds for which $^{13}$CO emission was detected as well.
Among the $^{13}$CO detections is cloud M33\_18, beyond the R$_{25}$ radius (see Fig. 3 as well).  
This is the first such detection and $^{13}$CO is detected in both transitions.
It is also of interest that the line widths decrease significantly from the inner regions (bottom of figure) to the outermost parts (top).  The narrow lines show that we are only observing a single cloud, allowing us to try to use the Virial theorem in the next section to estimate the cloud masses.   
Cloud 4 \citep[the]["Lonely Cloud"]{Gardan07} is very well-detected but is in an interarm region with little or no detected star formation despite the strong CO emission.  Clouds 4 and 18 have been observed with the Plateau de Bure Interferometer and will be discussed at length in a dedicated article.

While there is a general link between CO emission and HI and FIR ({\it e.g.} 70$\mu$m) emission, the CO emission of clouds 4 and 18 appears surprising even with the benefit of hindsight.  The 70$\mu$m emission, for example, 
is only very marginally detected at the positions of these strong CO detections.
However, something must provoke the condensation of atomic gas clouds to form H$_2$.
In the case of cloud 18, the HI spectra are complicated so perhaps the H$_2$ is the result of 
merging HI clouds \citep{Brouillet92,Ballesteros99,Hennebelle99,Heitsch05}.  In the other molecular clouds 
in the outer regions of M~33, this seems less likely because the HI spectra are not wider or more complicated 
than in many places where CO is not found.

\begin{figure}[t]
\begin{flushleft}
\includegraphics[angle=0,width=8.8cm]{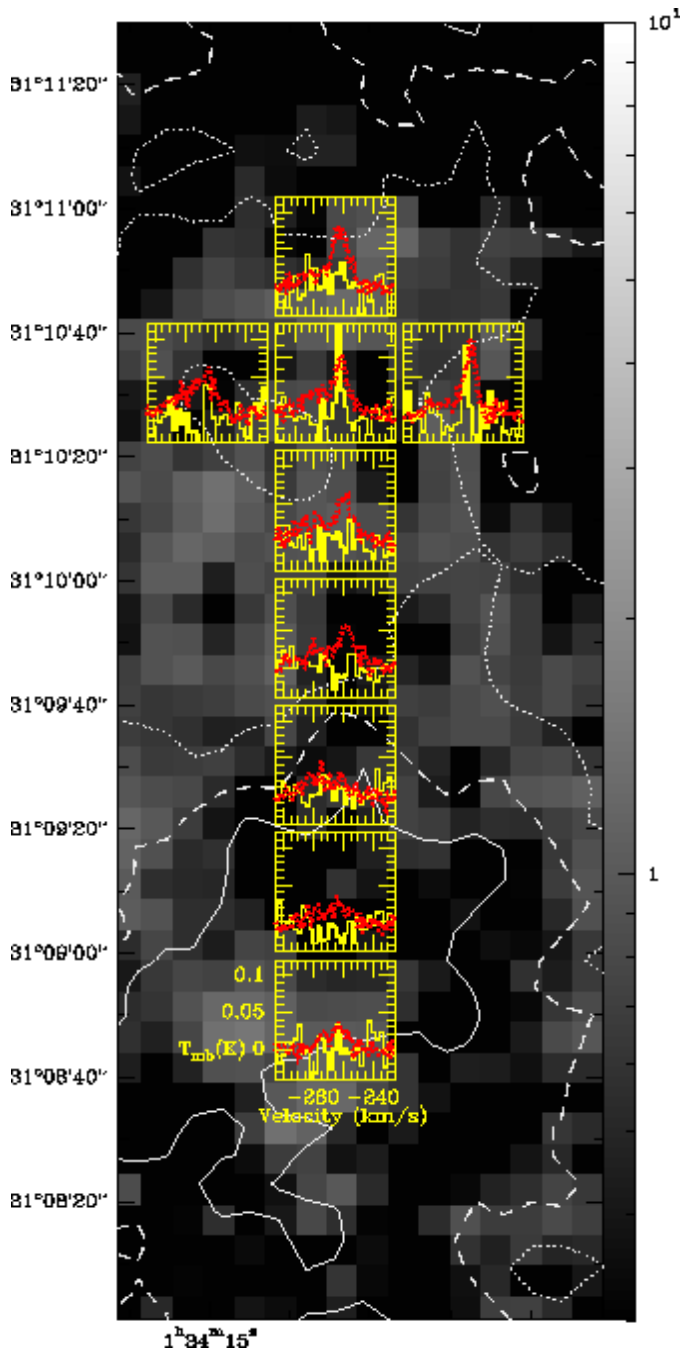}
\caption{ CO(1--0) spectra (yellow) and HI spectra (red dotted) of source M33\_18 and major offsets
on 70$\mu m$ image with HI contours at 0.8, 1.2 (dashed), and 1.6 $\times 10^{21}$ cm$^{-2}$ (dotted).  
The wedge shows the 70$\mu m$ image scale in MJy sr$^{-1}$.  The CO maximum is actually at the (-10.5,0) position 
shown in Fig. 6.}
\end{flushleft}
\end{figure}

\begin{figure}[t]
\begin{flushleft}
\includegraphics[angle=0,width=8.8cm]{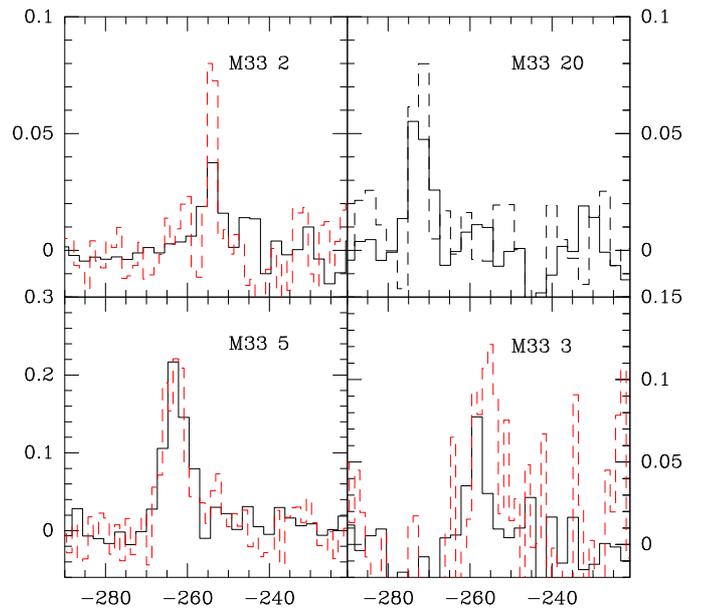}
\caption{ Spectra of the 4 clouds where only $^{12}$CO emission was detected.  The $J = 1 \rightarrow 0$
transition is in black and the $J = 2 \rightarrow 1$ is in red (dashed).  The spectra of M33\_20 show the offset 
(150,0) with a solid line and the dashed line is the (120,0) offset, both in CO(1--0).
The $x$-axis shows the velocity in km s$^{-1}$ and the $y$-axis the antenna temperature on the main beam scale.   
The order is in increasing declination from bottom to top, corresponding to increasing galactocentric distance.  }
\end{flushleft}
\end{figure}

\begin{figure}[t]
\begin{flushleft}
\includegraphics[angle=0,width=8.8cm]{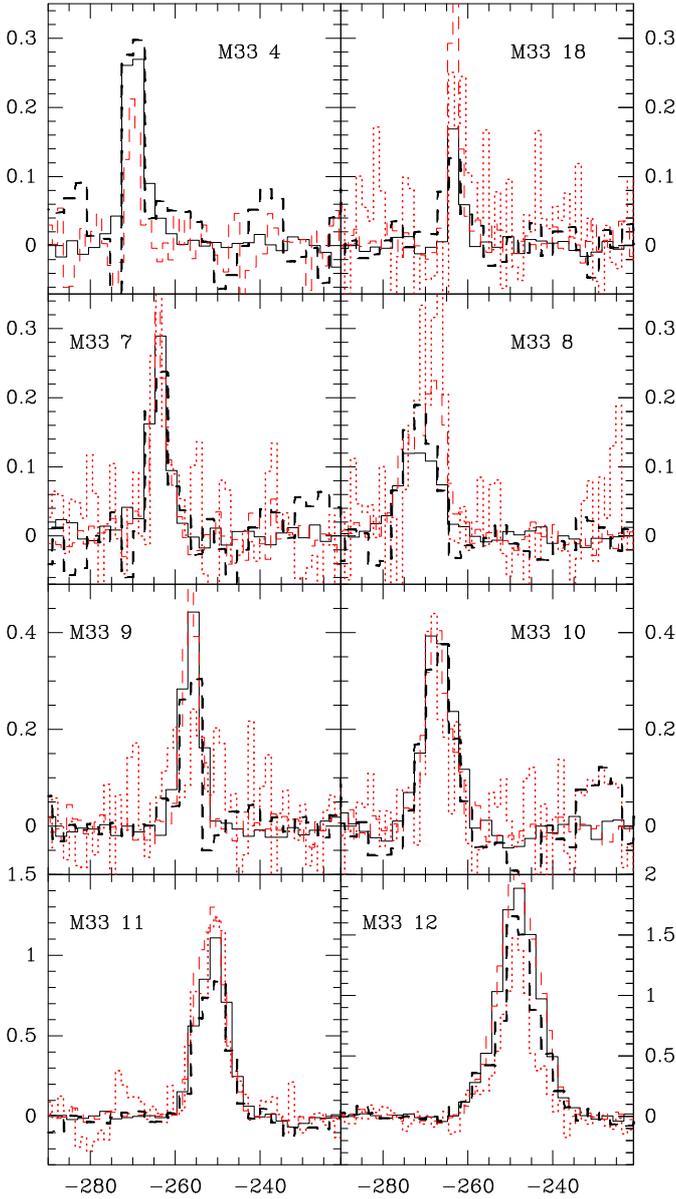}
\caption{ Spectra of the 8 clouds where $^{13}$CO emission was detected.  The $J = 1 \rightarrow 0$
transition is in black and the $J = 2 \rightarrow 1$ is in red.  The $^{13}$CO antenna temperatures are multiplied by 10.  The dotted red line is the $^{13}$CO(2--1) and the dashed black line is the $^{13}$CO(1--0) spectrum.
The $x$-axis shows the velocity in km$s^{-1}$ and the $y$-axis the antenna temperature on the main beam scale.  
The order is in increasing declination from bottom to top, corresponding to increasing galactocentric distance. 
Cloud 8 is most likely a blend of two clouds at different velocities \citep{Gardan07b} and in the spectra above
there is a clear difference between the CO(1--0) line shape and center compared with the (2--1) transition.}
\end{flushleft}
\end{figure}

\section{Determining the H$_2$ mass}

\subsection{via the Virial theorem applied to individual clouds}

Historically, cloud masses have been estimated assuming that molecular 
clouds are bound under their own gravity.  Galactic GMCs show a relationship between 
size and linewidth \citep{Larson81} which is difficult to explain in other ways.
The masses calculated are usually referred to as "Virial masses", twice as large 
as marginally gravitationally bound clouds.  In addition to GMCs, diffuse clouds are also observed
in the Galaxy \citep{Polk88} and these are probably unbound.  The scales we observe in M~33 correspond
to GMCs, justifying the use of the Virial method.  Since the only force opposing
gravity in the standard calculations is the internal motion, measured by the line width,
magnetic support could increase masses further if clouds really were stable long-lived 
"Virialized" objects.  Since the idea of virialized clouds came out, it has become clear
that GMCs are not long-lived entities but rather have "lifetimes" of order $10^7$ years
\citep[e.g.][]{Kawamura09}.  However, the "Virial" masses have generally
been found to be consistent with other means of measuring GMC masses, such as dust
measurements \citep{Sodroski95} or gamma-ray observations \citep[see e.g.][]{Combes91}.
Using the expression from \citet{Dickman86}, 

$$ M_{\rm vir} \approx {{\Delta V^2 D}\over {2 \gamma G}} $$

where $\Delta V$ is the half-power velocity width and $D$ the cloud diameter ($\sqrt{4 A/\pi}$ 
for other shapes).  The  $\gamma$ parameter is principally related to the broadening of the 
$^{12}$CO line due to its high optical depth \citep[see ][for details]{Dickman86}. 
Using this, the cloud masses (M$_{\rm vir}$) are given in 
Table 4, along with cloud velocities, line widths, and masses calculated using 
a $\ratio = 2 \times 10^{20} \Xunit$ factor (M$_{\rm lum}$). 

The Virial masses given are upper limits in the sense that we have assumed 
that the clouds are as big as the beam, 85 pc in diameter, whereas this is likely
an overestimate of the size of a single cloud \citep[e.g.][]{Solomon87}.  Furthermore,
multiple clouds would broaden the CO lines and again cause an overestimate of the mass.
However, the $\gamma=2$ value may be overestimated and is used for consistency 
with \citet{Braine_n4414b}.  These authors also found, through basic radiative transfer 
calculations and for modestly subsolar metallicites, that a low radiation field (as for the clouds in 
the outer disk of M~33) could counteract
the deeper CO photo-dissociation expected with decreasing metallicity, such that 
the CO-emitting part of a cloud was not necessarily smaller in an outer disk environment.
Nonetheless, the virial masses
given here are higher (possibly due to the large assumed size) in most cases
than galactic GMCs of similar line width \citep[using M$_{\rm vir} \approx 365 \Delta V^4 M_\odot$, eq. 6a
from ][]{Solomon87}.  Attributing an uncertainty is difficult; the sources of error are chiefly the size,
the $\gamma$ parameter, and the hypothesis of "Virial" equilibrium (gravity alone).  The size and $\gamma$ likely lead to an uncertainty of a factor 2 at most.  The presence of CO-bright isolated clouds in the outer disk 
is evidence that they are robust, and thus gravitationally bound, structures, comforting the Virial hypothesis.
If the clouds we identify were simply due to the fluctuating turbulent field then we would expect to find many weak lines for a single strong one (if indeed strong lines could be formed this way) and the \citet{Gardan07} observations suggest this is not the case.  The relatively strong $^{13}$CO emission shows that the $^{12}$CO line is highly optically thick, also arguing for a non-transient nature.
The spectra of some clouds, particularly in the inner disk, are complicated and single-gaussian fits lead to significant uncertainties in the linewidth and thus Virial mass.  These are a minority, however, so we estimate an uncertainty
of at most a factor 2 on the value given below.

With the exception of the two bright inner disk clouds, M33\_11 and M33\_12, 
which are consistent with a galactic $\ratio$ factor,
and the very uncertain sources, the virial masses are consistent with a 
$\ratio \approx 5 \times 10^{20} \Xunit$ factor (the average for clouds 4, 5, 7, 9, 10, and 18).

\begin{flushleft}
\tablecaption{Positions observed in CO.}
\tablefirsthead{
\hline
\hline
Name & RA & Dec &  $\delta$RA & $\delta$Dec  \\
 & & &  $(\arcsec)$ & $(\arcsec)$ \\
\hline
}
\tablehead{
\multicolumn{5}{l}{\small\sl continued from previous page}\\
\hline
Name & RA & Dec &  $\delta$RA & $\delta$Dec  \\
 & & & $(\arcsec)$ & $(\arcsec)$ \\
\hline
}
\tabletail{
\hline
\multicolumn{5}{r}{\small\sl continued on following page}\\
}
\tablelasttail{
\hline
\hline
}
\small
\begin{supertabular*}{88mm}{@{\extracolsep{\fill}}l rrrr}

M33\_1 & 01:34:17.67 &  31:05:12.0 &   0.0 &   0.0 \\
   M33\_2 & 01:34:06.13 &  31:03:09.7 &   0.0 &   0.0  \\
   &  &  & -20.5 &  -0.0 \\
   &  &  &  -0.0 & -20.5  \\
   &  &  &  20.5 &  -0.0  \\
   &  &  &  -0.0 &  20.5  \\
   M33\_3 & 01:34:34.86 &  31:00:42.9 &   0.0 &   0.0 \\
   &  &  & -20.5 &   0.0 \\
   &  &  &   0.0 & -20.5  \\
   M33\_4 & 01:34:16.81 &  30:59:30.0 & -10.5 & -10.5  \\
   &  &  & -10.5 &  -0.0  \\
   &  &  & -10.5 &  10.5  \\
   &  &  &   0.0 & -10.5  \\
   &  &  &   0.0 &   0.0  \\
   &  &  &   0.0 &  10.5  \\
   &  &  &  10.5 & -10.5  \\
   &  &  &  10.5 &   0.0  \\
   &  &  &  10.5 &  10.5  \\
   &  &  & -21.0 &  -0.0  \\
   &  &  &  21.0 &   0.0  \\
   &  &  &   0.0 & -21.0  \\
   &  &  &  -0.0 &  21.0  \\
   M33\_5 & 01:34:03.06 &  30:58:05.2 &   0.0 &   0.0  \\
   &  &  & -20.5 &   0.0  \\
   &  &  &   0.0 & -20.5  \\
   &  &  &  20.5 &  -0.0  \\
   &  &  &  -0.0 &  20.5  \\
   M33\_7 & 01:34:22.37 &  30:57:03.1 &   0.0 &   0.0  \\
   &  &  & -20.5 &  -0.0  \\
   &  &  &  -0.0 & -20.5  \\
   &  &  &  20.5 &  -0.0  \\
   &  &  &  -0.0 &  20.5  \\
   M33\_8 & 01:34:29.94 &  30:57:10.1 & -20.5 &  -0.0  \\
   &  &  &  20.5 &   0.0  \\
   &  &  &   0.0 &  20.5  \\
   &  &  &   0.0 & -20.5  \\
   &  &  &   0.0 &   0.0  \\
   M33\_9 & 01:34:03.72 &  30:55:01.8 & -20.5 & -20.5  \\
   &  &  & -20.5 &  -0.0  \\
   &  &  & -20.5 &  20.5  \\
   &  &  &  20.5 & -20.5  \\
   &  &  &  20.5 &   0.0  \\
   &  &  &  20.5 &  20.5  \\
   &  &  &  -0.0 &  20.5  \\
   &  &  &  -0.0 & -20.5  \\
   &  &  &   0.0 &   0.0  \\
  M33\_10 & 01:34:16.23 &  30:52:08.4 &   3.7 &  10.5  \\
   &  &  &   0.0 &   0.0  \\
  M33\_11 & 01:33:59.46 &  30:49:12.1 &  -4.6 & -28.6  \\
   &  &  &   0.0 &   0.0  \\
  M33\_12 & 01:34:09.30 &  30:49:06.1 &   0.0 &   0.0  \\
  M33\_18 & 01:34:13.31 &  31:10:32.0 &   0.0 &   0.0  \\
   &  &  & -20.5 &   0.0  \\
   &  &  &  -0.0 & -20.5  \\
   &  &  &  20.5 &   0.0  \\
   &  &  &   0.0 &  20.5  \\
   &  &  &  -0.0 & -10.5  \\
   &  &  & -10.5 &  -0.0  \\
   &  &  &   0.0 & -41.0  \\
   &  &  &  -0.0 & -61.5  \\
   &  &  &  -0.0 & -82.0  \\
   &  &  &  -0.0 &-102.5  \\
  M33\_20 & 01:34:39.79 &  31:08:23.8 &   0.0 &   0.0  \\
   &  &  &  30.0 &  -0.0  \\
   &  &  &  60.0 &   0.0  \\
   &  &  &  90.0 &  -0.0  \\
   &  &  & 120.0 &   0.0  \\
   &  &  & 150.0 &  -0.0  \\
   &  &  & 180.0 &   0.0  \\
   &  &  & 210.0 &   0.0  \\
   &  &  & 240.0 &   0.0  \\
  M33\_22 & 01:34:54.09 &  31:13:25.5 &   0.0 &   0.0  \\
  M33\_23 & 01:34:45.28 &  31:12:57.5 &   0.0 &   0.0  \\
   &  &  &  22.6 &   5.6  \\
   &  &  &  45.2 &  11.2  \\
   &  &  &  67.8 &  16.8  \\
   &  &  &  90.4 &  22.4  \\
   &  &  & -10.5 &   0.0  \\
  M33\_24 & 01:34:52.39 &  31:08:36.0 &   0.0 &   0.0  \\
  M33\_25 & 01:34:22.69 &  31:11:54.4 &   0.0 &   0.0  \\
   &  &  &  -0.0 & -10.5  \\
  M33\_26 & 01:34:40.88 &  31:13:40.0 &   0.0 &   0.0  \\
  M33\_27 & 01:34:51.85 &  31:09:39.6 &   0.0 &   0.0  \\
  M33\_34 & 01:34:57.40 &  31:05:13.9 &   0.0 &   0.0  \\
\hline
\end{supertabular*}
\end{flushleft}

\subsection{From isotopic lines and ratios}

Another approach to estimating cloud masses is to measure isotopic lines 
assumed to be optically thin.  In principle, this allows one to "count" the 
molecules of the optically thin species. In practice, this is quite difficult but 
isotopic line ratios provide a means to 
determine whether the GMCs are intrinsically very different from Galactic clouds.  
Eight of the clouds have been observed in $^{13}$CO and all of them were 
detected in the $^{13}$CO(1--0) line and all but one in $^{13}$CO(2--1).

Assuming optically thin $^{13}$CO(1--0) emission and Local Thermodynamic 
Equilibrium at a temperature $T_{\rm ex} >> T_{\rm CMB}$, the remarkably 
narrow range in $^{12/13}$CO line ratios (Table 5) yields $\ratio$ values within a factor 2
of $10^{20} \Xunit$ \citep[see formulae 14.40 and 14.41 in ][]{Rohlfs04}
for realistic excitation temperatures (assumed 10 -- 30 K).  This is true for both
the $^{13}$CO($J=1-0$) and $J=2-1$ transitions.
The calculations assumed a $^{13}$CO abundance with respect to H$_2$ of
$(1.2 \times 10^6)^{-1}$ -- lower $^{13}$CO abundances will yield a higher 
$\ratio$ value. Of course, a fraction of the $^{13}$CO is probably not optically thin and the 
departure from LTE will be much greater in $^{13}$CO than $^{12}$CO
due to the inefficient radiative excitation.  Furthermore, the fraction of the GMC
where $^{13}$CO is dissociated will be much greater than for $^{12}$CO.
The combination of these ingredients likely explains the much lower $\ratio$
value obtained in this way.  Low $^{12/13}$CO line ratios lead to high $\ratio$ ratios.

The $^{13}$CO integrated intensities and $^{12/13}$CO line ratios for each 
transition are given in Table 5.  
The ratios are of order 10 for the $J=2-1$ transition and slightly higher for 
the $J=1-0$.  These values are typical of (large) galaxies at large scales \citep{Rickard85,Sage91,Paglione01}.  Values for individual clouds in the Milky Way are usually lower, in the range 3 -- 5 \citep{Polk88,Digel94}
but may not include all of the low column density material
with significant $^{12}$CO but little $^{13}$CO emission.  Since M33 has a metallicity 
which is subsolar by a factor 2, the $^{12/13}$CO line ratio could be expected to be
higher, as the optically thick $^{12}$CO will be less affected than the $^{13}$CO by
the lower abundance.

The $^{13}$CO spectra shown in Fig. 6 are multiplied by 10 and can be seen to lay 
well over the $^{12}$CO spectra.  The line ratios observed in M33 are thus 
in good agreement with the picture in which molecular clouds in M33 are similar to
their Galactic counterparts but with a lower metallicity and a $\ratio$
higher by about a factor two than in the Milky Way disk.  

\begin{table}[htbp]
\caption{CO(1--0) detections.}
\begin{tabular*}{88mm}{@{\extracolsep{\fill}}l rrrr}
\hline
Name & $\delta$RA & $\delta$Dec & I$_{\rm CO(1-0)}$ & $rms$ \\
 &  $(\arcsec)$ & $(\arcsec)$ & (K km s$^{-1}$) & (mK) \\
\hline
 M33\_2 &  0.0 &  0.0 &  0.20 $\pm$ 0.03 &  7.1 \\
 M33\_3 &  0.0 &  0.0 &  0.42 $\pm$ 0.06 & 17.5 \\
 M33\_4 &-10.5 &-10.5 &  1.20 $\pm$ 0.09 & 21.1 \\
 M33\_4 &-10.5 & 0.0 &  1.50 $\pm$ 0.08 & 19.1 \\
 M33\_4 &-10.5 & 10.5 &  0.99 $\pm$ 0.13 & 21.5 \\
 M33\_4 &  0.0 &-10.5 &  1.20 $\pm$ 0.06 & 17.7 \\
 M33\_4 &  0.0 &  0.0 &  1.83 $\pm$ 0.05 & 12.5 \\
 M33\_4 &  0.0 & 10.5 &  1.25 $\pm$ 0.08 & 16.4 \\
 M33\_4 & 10.5 &-10.5 &  0.64 $\pm$ 0.06 & 19.3 \\
 M33\_4 & 10.5 &  0.0 &  0.79 $\pm$ 0.06 & 16.9 \\
 M33\_4 & 10.5 & 10.5 &  0.60 $\pm$ 0.08 & 17.9 \\
 M33\_4 &-21.0 & 0.0 &  0.99 $\pm$ 0.07 & 17.3 \\
 M33\_4 & 0.0 & 21.0 &  0.94 $\pm$ 0.06 & 15.9 \\
 M33\_5 &  0.0 &  0.0 &  1.47 $\pm$ 0.06 & 14.9 \\
 M33\_5 &-20.5 &  0.0 &  0.24 $\pm$ 0.05 & 15.5 \\
 M33\_5 & 20.5 & 0.0 &  0.48 $\pm$ 0.05 & 16.2 \\
 M33\_5 & 0.0 & 20.5 &  0.42 $\pm$ 0.07 & 16.8 \\
 M33\_7 &  0.0 &  0.0 &  1.50 $\pm$ 0.04 & 12.5 \\
 M33\_7 &-20.5 & 0.0 &  0.36 $\pm$ 0.07 & 17.9 \\
 M33\_7 & 20.5 & 0.0 &  0.41 $\pm$ 0.07 & 16.4 \\
 M33\_7 & 0.0 & 20.5 &  1.04 $\pm$ 0.06 & 16.5 \\
 M33\_8 &-20.5 & 0.0 &  0.67 $\pm$ 0.08 & 18.3 \\
 M33\_8 &  0.0 & 20.5 &  0.75 $\pm$ 0.07 & 18.0 \\
 M33\_8 &  0.0 &-20.5 &  0.35 $\pm$ 0.08 & 16.5 \\
 M33\_8 &  0.0 &  0.0 &  1.44 $\pm$ 0.04 &  8.1 \\
 M33\_9 &-20.5 &-20.5 &  0.53 $\pm$ 0.08 & 25.0 \\
 M33\_9 & 20.5 &-20.5 &  0.88 $\pm$ 0.10 & 29.6 \\
 M33\_9 & 20.5 &  0.0 &  0.59 $\pm$ 0.07 & 20.6 \\
 M33\_9 & 0.0 & 20.5 &  1.25 $\pm$ 0.08 & 22.1 \\
 M33\_9 & 0.0 &-20.5 &  0.74 $\pm$ 0.15 & 21.8 \\
 M33\_9 &  0.0 &  0.0 &  2.51 $\pm$ 0.04 & 11.7 \\
M33\_10 &  3.7 & 10.5 &  3.46 $\pm$ 0.14 & 28.9 \\
M33\_10 &  0.0 &  0.0 &  3.55 $\pm$ 0.08 & 18.4 \\
M33\_11 & -4.6 &-28.6 &  9.99 $\pm$ 0.17 & 33.5 \\
M33\_11 &  0.0 &  0.0 &  9.73 $\pm$ 0.08 & 16.9 \\
M33\_12 &  0.0 &  0.0 & 22.78 $\pm$ 0.10 & 18.9 \\
M33\_18 &  0.0 &  0.0 &  0.52 $\pm$ 0.06 & 19.5 \\
M33\_18 &-20.5 &  0.0 &  0.26 $\pm$ 0.06 & 19.8 \\
M33\_18 & 0.0 &-10.5 &  0.24 $\pm$ 0.03 &  9.8 \\
M33\_18 &-10.5 & 0.0 &  0.61 $\pm$ 0.02 &  7.6 \\
M33\_20 &120.0 &  0.0 &  0.26 $\pm$ 0.03 & 11.4 \\
M33\_20 &150.0 & 0.0 &  0.33 $\pm$ 0.04 & 10.2 \\
\hline
\end{tabular*}
\end{table}
 
\begin{table}[htbp]
\caption{CO(2--1) detections.}
\begin{tabular*}{88mm}{@{\extracolsep{\fill}}l rrrr}
\hline
Name & $\delta$RA & $\delta$Dec & I$_{\rm CO(2-1)}$ & $rms$ \\
 & $(\arcsec)$ & $(\arcsec)$ & (K km s$^{-1}$) & (mK) \\
\hline
 M33\_2 &  0.0 &  0.0 &  0.24 $\pm$ 0.02 & 11.6 \\
 M33\_3 &  0.0 &  0.0 &  0.67 $\pm$ 0.12 & 40.8 \\
 M33\_4 &  0.0 &  0.0 &  0.57 $\pm$ 0.06 & 31.2 \\
 M33\_4 &-10.5 &-10.5 &  1.04 $\pm$ 0.11 & 40.5 \\
 M33\_4 &-10.5 & 0.0 &  1.09 $\pm$ 0.11 & 42.4 \\
 M33\_4 &  0.0 &-10.5 &  0.81 $\pm$ 0.08 & 40.2 \\
 M33\_4 &  0.0 &  0.0 &  1.31 $\pm$ 0.04 & 19.7 \\
 M33\_4 &  0.0 & 10.5 &  0.56 $\pm$ 0.10 & 33.5 \\
 M33\_4 & 10.5 & 10.5 &  0.30 $\pm$ 0.06 & 41.0 \\
 M33\_4 &-21.0 & 0.0 &  0.57 $\pm$ 0.10 & 37.5 \\
 M33\_4 & 0.0 & 21.0 &  0.84 $\pm$ 0.11 & 41.5 \\
 M33\_5 &  0.0 &  0.0 &  1.38 $\pm$ 0.09 & 32.0 \\
 M33\_5 & 0.0 & 20.5 &  0.37 $\pm$ 0.09 & 34.9 \\
 M33\_7 &  0.0 &  0.0 &  1.40 $\pm$ 0.05 & 21.5 \\
 M33\_7 & 0.0 & 20.5 &  0.83 $\pm$ 0.08 & 34.2 \\
 M33\_8 &  0.0 & 20.5 &  0.97 $\pm$ 0.09 & 40.9 \\
 M33\_8 &  0.0 &  0.0 &  2.28 $\pm$ 0.06 & 14.8 \\
 M33\_9 & 0.0 & 20.5 &  0.85 $\pm$ 0.08 & 35.4 \\
 M33\_9 &  0.0 &  0.0 &  2.34 $\pm$ 0.05 & 19.6 \\
 M33\_9 & 20.5 &-20.5 &  1.09 $\pm$ 0.09 & 40.2 \\
M33\_10 &  3.7 & 10.5 &  2.84 $\pm$ 0.09 & 27.1 \\
M33\_10 &  0.0 &  0.0 &  2.99 $\pm$ 0.06 & 18.7 \\
M33\_11 & -4.6 &-28.6 &  6.86 $\pm$ 0.09 & 25.6 \\
M33\_11 &  0.0 &  0.0 & 10.72 $\pm$ 0.07 & 21.0 \\
M33\_12 &  0.0 &  0.0 & 23.76 $\pm$ 0.10 & 25.7 \\
M33\_18 &  0.0 &  0.0 &  0.40 $\pm$ 0.05 & 31.4 \\
M33\_18 &-10.5 & 0.0 &  1.22 $\pm$ 0.03 & 17.6 \\
 \hline
\end{tabular*}
\end{table}

\begin{table}[htbp]
\caption{Results of gaussian fits to the detected clouds in CO(1--0).
 }
\begin{tabular*}{88mm}{@{\extracolsep{\fill}}l rrrr}
\hline
Name & velocity & line width & M$_{\rm vir}$ & M$_{\rm lum}$ \\
 & (km s$^{-1}$) & (km s$^{-1}$) & (1000 M$_\odot$) & (1000 M$_\odot$) \\
\hline
  M33\_2 & -254.1 $\pm$ 0.5 &  5.0 $\pm$ 1.2 &  125 $\pm$ 60 &    6 \\
  M33\_9 & -256.2 $\pm$ 0.1 &  5.3 $\pm$ 0.2 &  140 $\pm$ 9 &   81 \\
 M33\_11 & -251.2 $\pm$ 0.1 &  8.5 $\pm$ 0.1 &  361 $\pm$ 10 &  312 \\
 M33\_12 & -248.2 $\pm$ 0.0 & 11.4 $\pm$ 0.1 &  643 $\pm$ 11 &  730 \\
  M33\_3 & -258.7 $\pm$ 0.4 &  5.5 $\pm$ 1.2 &  150 $\pm$  66 &   20 \\
  M33\_4 & -269.8 $\pm$ 0.1 &  4.4 $\pm$ 0.2 &   95 $\pm$ 7 &   58 \\
  M33\_5 & -263.0 $\pm$ 0.1 &  6.7 $\pm$ 0.3 &  221 $\pm$ 23 &   55 \\
  M33\_7 & -263.9 $\pm$ 0.1 &  4.1 $\pm$ 0.2 &   82 $\pm$ 9 &   48 \\
  M33\_8 & -271.6 $\pm$ 0.2 & 10.7 $\pm$ 0.4 &  567 $\pm$ 40 &   44 \\
 M33\_10 & -267.1 $\pm$ 0.1 &  7.6 $\pm$ 0.2 &  284 $\pm$ 17 &  105 \\
 M33\_18 & -263.3 $\pm$ 0.1 &  2.5 $\pm$ 0.2 &   31 $\pm$ 4 &   19 \\
  \hline
\end{tabular*}
Velocities and line widths, with the errors estimated by the gaussian fitting 
algorithm in CLASS are provided in columns 2 \& 3.  Making the assumption 
that the cloud radius is equal to the beamsize, virial masses are calculated as in 
\citet{Braine_n4414b} and the errors using only the uncertainty in $\Delta V$.
Masses from the CO(1--0) line intensity are given in column 5 using the line 
area from Table 2a and a "standard" galactic conversion $\ratio = 2 \times 10^{20} \Xunit $.
The first 4 clouds have gaussian fits made using the 2.6 km s$^{-1}$ channel width whereas 
the latter 7 fits have been made using data at 0.52 km s$^{-1}$ channel width.  Cloud 8 
appears to have two velocity components, producing the broad line.  
Cloud 2 has a large error associated with the CO(1--0) observation; 
the gaussian fit to the CO(2--1) line yields a width of $2.2\pm0.44 kms^{-1}$
but the CO(2--1) beam is much smaller and may not include all the emission.
\end{table}

\begin{table}[htbp]
\caption{Table 5.  $^{13}$CO observations}
\begin{tabular*}{88mm}{@{\extracolsep{\fill}}l rrrrr}
\hline
Name & I$_{\rm (1-0)}$ & $\sigma_{1-0}$   & I$_{\rm (2-1)}$ & I$\left(\frac{12}{13}\right)_{(1-0)}$ & 
  I$\left(\frac{12}{13}\right)_{(2-1)}$    \\
  & Kkm$s^{-1}$ & mK & Kkm$s^{-1}$ & & \\
\hline
 M33\_4 &  0.15 $\pm$ 0.02 & 5.6
 &31.2 mK &  12.2 &   --- \\
 M33\_7 &  0.14 $\pm$ 0.02 & 4.2
 &  0.16 $\pm$ 0.01 &  10.7 &   8.5 \\
 M33\_8 &  0.13 $\pm$ 0.02 & 3.3
 &  0.23 $\pm$ 0.02 &  11.2 &  10.0 \\
 M33\_9 &  0.16 $\pm$ 0.01 & 3.9
 &  0.23 $\pm$ 0.04 &  15.7 &  10.1 \\
M33\_10 &  0.29 $\pm$ 0.02 & 5.2
 &  0.22 $\pm$ 0.02 &  12.2 &  13.8 \\
M33\_11 &  0.77 $\pm$ 0.02 & 5.1
 &  1.09 $\pm$ 0.04 &  12.6 &   9.8 \\
M33\_12 &  1.78 $\pm$ 0.02 & 4.6
 &  1.21 $\pm$ 0.03 &  12.8 &  19.7 \\
M33\_18 &  0.07 $\pm$ 0.01 & 2.9
 &  0.11 $\pm$ 0.02 &   8.9 &  11.4 \\
  \hline
\end{tabular*}
\end{table}

\section{Conclusions}

We present the first detections of both $^{12}$CO and  $^{13}$CO in the extreme outer disk ($R > R_{25}$) of a subsolar metallicity galaxy.  Several clouds have been detected in our search for molecular gas in the outer disk, enabling us to try to identify the environments in which detectable quantities of H$_2$ are found in the dim outer 
disks of spirals.  
The HI antenna temperature and column density and the level of star formation 
as traced by FIR or H$\alpha$ emission are indicators of the likelihood of detecting CO but quite imperfect 
because two of the strongest outer disk clouds would not have been found that way.
In and of itself, the unreliability of this link is interesting.  

As we are close to spatially resolving clouds, the "Virial" method is used to estimate molecular cloud masses and thus 
the $\ratioo$ ratio.   The inner disk clouds measured are compatible with a Galactic $\ratioo$ factor.  Our estimate of $\ratio$ in the outer disk of M~33 is $5 \times 10^{20} \Xunit$ with an uncertainty of at most a factor 2.  The $^{12/13}$CO line ratios are compatible with a $\ratio$ factor somewhat greater than in the Milky Way molecular ring, such as the $\ratioo \sim 5 \times 10^{20} \Xunit$ estimated by the Virial method presented here.
Adopting this conversion factor results in a lower star formation efficiency (SFR / M$_{\Htwo}$) than previously found but the SFE remains higher than the SFE in large local universe spirals \citep[e.g. ][]{Kennicutt98b}.
 
\bibliographystyle{aa}
\bibliography{jb}

\end{document}